\documentclass[11pt,a4paper,serif]{article}
\usepackage[margin=2.25cm]{geometry}
\usepackage[utf8]{inputenc}
\usepackage{graphicx}
\usepackage{natbib}
\usepackage[bottom]{footmisc}
\usepackage{setspace}
\usepackage{booktabs,tabularx}
\usepackage{array}
\usepackage{lscape}
\usepackage[toc,page]{appendix}
\usepackage{pgfplots}
\pgfplotsset{width=10cm,compat=1.5}
\usepackage{csquotes}
\usepackage{tikz}
\usepackage{pgf-pie}
\usepackage{float}

\usepackage{eurosym}
\usepackage{lscape}
\usepackage[labelsep=period]{caption,subcaption}
\usepackage{multirow}
\usepackage{amsmath}
\usepackage{pgfplotstable}
\usepackage[font=scriptsize]{caption}
\usepackage[affil-it]{authblk}
\usepackage{url}

\pgfplotsset{compat=1.10}
\usepgfplotslibrary{statistics}
\usepackage{color,xcolor,soul,hyperref}
\hypersetup{colorlinks=true,citecolor=blue,filecolor=black,linkcolor=blue,urlcolor=black}

\begin{document}

	% Title
	\title{The Time Function of Stock Price: An integral white noise model and its time- and frequency- domain characteristics}
	
	%\author{}
	% Authors
	\author{Shengfeng Mei%
		\thanks{E-mail: shengfeng.mei@glasgow.ac.uk, corresponding author}}
	\affil{Adam Smith Business School, University of Glasgow}
	
	\author{Hong Gao%
		}
	\affil{Tsinghua University}

	\date{}
	
	\maketitle
	\pagenumbering{arabic}
	%\thispagestyle{empty}
	
	% Abstract
	\begin{abstract}
		
		\noindent  This paper defines the quantitative relationship between the stock price and time as a time function. Based on the empirical evidence that “the log-return of a stock is the series of white noise”, a mathematical model of the integral white noise is established to describe the phenomenon of stock price movement. A deductive approach is used to derive the auto-correlation function, displacement formula and power spectral density (PSD) of the stock price movement, which reveals not only the characteristics and rules of the movement but also the predictability of the stock price. The deductive fundamental is provided for the price analysis, prediction and risk management of portfolio investment.
	\end{abstract}
	

	%	\tableofcontents
	\clearpage
	
	\section{Introduction}\label{sec:introduction}
As early as 1900, a French mathematician who is the founder of quantitative finance, Louis Bachelier’s PhD thesis, ``The Theory of Speculation”, derives a probabilistic method to study the time-varying rule of the stock price \citep{courtault2000louis,jovanovic2012bachelier,weatherall2018peculiar}. Bachelier discovers that the change in the stock price is completely stochastic and defines the price at each time point as a random variable. Then, he establishes a Brownian model to describe the stock price movement. Later in 1958, a high-energy physicist from the Naval Research Laboratory (NRL) found that the stock price Brownian model may have negative values and developed it into the geometric Brownian model \citep{osborne1959brownian}. Owing to the assumption of the random variable, the geometric Brownian model also shares a zero mean and fails to explain the long-run linear trend existing in the stock price movement. \cite{samuelson2015rational} embeds a linear drift in the model to cover this. A geometric Brownian model with the drift part is founded \citep{szpiro2011pricing}.

With the development of the mathematical model of stock price \citep{osborne1959brownian,samuelson2015rational}, a paradigm of quantitative finance is produced. The paradigm states that the quantitative relationship between stock prices and time can be abstracted as a random variable. Meanwhile, it is believed use the probabilistic approach to study how the stock price will change as time changes. The definition of a random variable states that it is a real function defined in a sample space rather than a function with respect to time variables. In contrast, the stock price and time relationship are commonly considered random variables. As a result, it is mistakenly concluded that the variance of the stock price is proportional to time, breaching real life. Besides, the random variable captures all the sample functions in all states rather than a single sample function in only one state. The danger of misusing random variables will be discussed later.

This paper redefines the quantitative relationship between stock prices and time as a certain time function based on the observation that stock prices and time has one-to-one correspondence. It uses an analytical approach to investigate the time-varying process and rules of stock price and obtains the time- and frequency-domain characteristics of stock price movement.

\section{The Integral White Noise Model of Stock Price}
Let $s(t)$ be the price of the stock at time $t$. For each determined time $t$, there is a certain $s(t)$. Therefore, $s(t)$ is a deterministic function of time $t$.

\subsection{Log-Return of Stock}
Assume that $y(t)=\ln{s(t)}$ is the logarithmic stock price (or, stock price). Then, the log-return of stock is:
			\begin{eqnarray} 
				r(t) = y(t+\Delta t)-y(t) .
			\end{eqnarray}
Empirical analyses from literature \citep{working1934random,kendall1953analysis,osborne1959brownian,fama1995random,samuelson2016proof} demonstrate that the short-term log-return of stock prices is random, and the stock price is subject to a random walk with incremental white noise.

\subsection{Integral White Noise Model}
According to ``The short-term log-return of stock prices is white noise”, we make the following basic hypothesis (law): In the time interval $(0,+\infty)$ , the first order condition $\Delta y(t)$ of the stock price $y(t)$ at the minute time change $\Delta t$ is:
			\begin{eqnarray} \label{eq:2}
				\Delta y(t) = y(t+\Delta t)-y(t)=x(t) .
			\end{eqnarray}
where $x(t)$ is a white noise sample function with mean zero. The above equation can be regarded as a discretized differential equation. Let $y(0)=0$, and the stock price can be calculated as:
			\begin{eqnarray}\label{eq:3}
				y(t) =\int_{0}^{t} x(t)dt .
			\end{eqnarray}
Obviously, the stock price $y(t)$ is the variable-limit integral of the white noise sample function $x(t)$ and, therefore, the integral white noise model is non-linear and time-varying.

The equation (\ref{eq:3}), the white noise integral model, has the following characteristics: Firstly, it can accurately calculate historical data of $y(t)$ based on historical data of $x(t)$. Besides, the time domain and frequency domain characteristics of $y(t)$ in the past, present and future can be analyzed according to the time domain and frequency domain characteristics of $x(t)$. Therefore, the mathematical model can be used to describe and interpret the phenomenon, characteristics and laws of the stock price $y(t)$ fluctuation.

\subsection{Time- and Frequency-Domain Characteristics of White Noise}
In the basic assumption of equation (\ref{eq:2}), the white noise sample function $x(t)$ is defined as follows:
			\begin{eqnarray}\label{eq:4}
				\lim_{T \to \infty} \frac{1}{2T} \int_{-T}^{T} x(t)dt = 0 .
			\end{eqnarray}
			\begin{eqnarray}\label{eq:5}
				R_{x}(\tau)=N_0 \delta(t) .
			\end{eqnarray}
where $R_x (\cdot)$ denotes the auto-correlation function of $x(t)$. $N_0$ is a positive constant and $\delta(t)$ is a unit impulse function such that:
                \begin{eqnarray}
				\delta(t) = \begin{cases} 
                             +\infty, & t=0 \\
                             0,       & t \neq 0
                            \end{cases}, \nonumber \\ 
                \int_0^t \delta(t)dt = 1 . \nonumber            
			\end{eqnarray}
It is clear that $x(t)$ has autocorrelation if and only if the lag $\tau = 0$. In other words, $x(t)$ has no auto-correlation as the lag $\tau \neq 0$. Thus, the signal waveform of the white noise $x(t)$ in the time domain is a series of random pulses with infinitely narrow width and extremely fast fluctuations.

The white noise $x(t)$ is a wide-sense stationary (WSS) process. According to the Wiener-Khinchin theorem, the auto-correlation function $R_{x}(\tau)$ and the power spectral density (PSD) of $x(t)$ form a Fourier transform pair, giving the expression of PSD as:
                \begin{eqnarray}\label{eq:6}
				P_{x}(\omega)=N_0 ,
			\end{eqnarray}
where
\begin{eqnarray}
				\omega=2\pi \xi \nonumber.
			\end{eqnarray}
$\xi$ representing frequency (e.g. if time is measured in seconds, then frequency is in hertz) and, thus, $\omega$ represents the angular frequency. $N_0$ is a positive real constant, indicating that the PSD of white noise $x(t)$ has a uniform distribution throughout the frequency axis $(-\infty,+\infty)$. The physical meaning of $N_0$ represents the average power produced by the white noise signal on the unit resistance.

The definition of white noise above is merely defined in the time domain. The mean of the sample function is zero, and the PSD is uniformly distributed in the entire frequency axis $(-\infty,+\infty)$. Note that there is no probability distribution involving the white noise sample function. The distribution of $x(t)$ can have different forms, for example, it can have a Gaussian form and then the equation (\ref{eq:3}) is the Wiener process (Brownian motion).

The combination of equations (\ref{eq:4}) and (\ref{eq:5}) is an idealized mathematical model because its PSD is ``constant” and the autocorrelation function is an ``impact function”. Therefore, it has the advantages of simple processing and convenient calculating. It is an essential part of mathematical phenomenon study in the theoretical research.

\section{The System of Stock Price Generator}
Equation (\ref{eq:3}) shows that the stock price $y(t)$ is the integral of white noise sample $x(t)$ within the interval $[0,t]$. From the perspective of signal analysis and processing, stock price $y(t)$ is the output produced when the white noise signal $x(t)$ excites the nonlinear time-varying system. Figure (\ref{fig:my_label}) shows the system.

\begin{figure}[htbp]
    \centering
    
\tikzset{every picture/.style={line width=0.75pt}} %set default line width to 0.75pt        

\begin{tikzpicture}[x=0.75pt,y=0.75pt,yscale=-1,xscale=1]
%uncomment if require: \path (0,300); %set diagram left start at 0, and has height of 300

%Shape: Rectangle [id:dp19168831118290375] 
\draw   (350.33,127) -- (432,127) -- (432,167) -- (350.33,167) -- cycle ;
%Straight Lines [id:da37854849475789654] 
\draw    (432.33,149) -- (504.33,149) ;
%Straight Lines [id:da6980682247429904] 
\draw    (278.33,150) -- (350.33,150) ;
%Straight Lines [id:da7778771870138299] 
\draw    (151.33,150) -- (223.33,150) ;
%Straight Lines [id:da5303120017450353] 
\draw    (223.33,150) -- (256.33,108) ;
%Curve Lines [id:da6821660366343705] 
\draw    (234.33,122) .. controls (269.07,116.21) and (272.15,130.91) .. (277.72,148.12) ;
\draw [shift={(278.33,150)}, rotate = 251.57] [color={rgb, 255:red, 0; green, 0; blue, 0 }  ][line width=0.75]    (10.93,-3.29) .. controls (6.95,-1.4) and (3.31,-0.3) .. (0,0) .. controls (3.31,0.3) and (6.95,1.4) .. (10.93,3.29)   ;

% Text Node
\draw (116,140) node [anchor=north west][inner sep=0.75pt]   [align=left] {$\displaystyle x( t)$};
% Text Node
\draw (516,139) node [anchor=north west][inner sep=0.75pt]   [align=left] {$\displaystyle y( t)$};
% Text Node
\draw (244,154) node [anchor=north west][inner sep=0.75pt]   [align=left] {$\displaystyle K$};
% Text Node
\draw (381,128) node [anchor=north west][inner sep=0.75pt]   [align=left] {$\displaystyle \int $};

\end{tikzpicture}
    \caption{The System of Stock Price Generator.}
    \label{fig:my_label}
\end{figure}

Since the PSD of white noise $x(t)$ is constant, the PSD of output $y(t)$ completely depends on the transfer function of the system. So far, the study of the random walk of stock price can be equally transferred to the study of the characteristics of a certain system. The system shown contains two components: a switch and an integrator. The function of the switch is to cut off the white noise input signal $x(t)$ defined in the interval $(-\infty,+\infty)$ in order to obtain the sampling signal $x_k (t)$ defined in the interval $[0,t]$. Meanwhile, the function of the integrator is to perform an integral of $x_k (t)$ and yield the output. Besides, the integrator is the transfer function model of the system.

The mathematical model of the switch can be expressed as:
                \begin{eqnarray} \label{eq:7}
				K(t) = \begin{cases} 
                             1, & 0\leq t \leq T \\
                             0,       & otherwise
                            \end{cases}.       
			\end{eqnarray}
Apparently, $K(t)$ is a non-linear function and the closing process of the switch changes dynamically with time. Consequently, the sampling signal can be given by:
                \begin{eqnarray}\label{eq:8}
				x_k (t) = K(t) \cdot x(t) .
			\end{eqnarray}
As can be seen above, the sampled signal $x_k (t)$ is the product of the white noise signal, $x (t)$, and the switch, $K (t)$. The process of truncating the $x (t)$ into $x_k (t)$ is equivalent to adding a rectangular window function to $x (t)$. Owing that $K (t)$ is not possible to perform a full-cycle truncation of the harmonic components of all frequencies in the white noise signal $x (t)$, it will generate a rate leakage effect in the frequency domain and cause DC component in $x_k (t)$. A linear trend term will be formed in $y(t)$ as the integrator operates. 

The existence of the switch is based on a hypothesis that a listed company may run for infinitely long, and its stock may also exist forever. However, the time data of the stock price is finite, meaning that the available data begins when the shares are issued, i.e. at time $t=0$, and ends today, i.e. at time $t=T$. The finite data also convey that the stock price has a discrete time interval. If we arbitrarily define that the time interval of a stock is infinite, we allow the sustainability of the company and its stock. A sustainable company's equity is risk-free in the long run, while no financial asset is completely risk-free in reality, even government bonds. The financial market mainly consists of various risky assets.

The integrator in the system has low-pass filtering characteristics. It will amplify the low-frequency components in $x_k (t)$ and reduce the high-frequency components. Thus, the system output signal $y(t)$ mainly comprises slowly varying low-frequency components superimposed. Besides, the integrator is memory. So, the output of the current moment of the system is not only related to the input at present but also the input at all times before. Therefore, stock prices have ``memory” or ``relevance”.

\section{The Characteristics of Time Domain}
\subsection{Time Autocorrelation Function}
The autocorrelation function of stock price $y(t)$ is given by:
                \begin{eqnarray}\label{eq:9}
				R_y(\tau) &=& \overline{y(t-\tau)y(t)} \nonumber \\
                       \; &=& \int_0^{t-\tau} \int_0^{t} x(u)x(v)dudv \nonumber \\
                       \; &=& \int_0^{t-\tau} \int_0^{t} N_0 \delta(u-v)dudv \nonumber \\
                       \; &=& \int_0^{t-\tau} N_0 du \nonumber \\
                       \; &=& N_0 (t-\tau)
			\end{eqnarray}
where $\tau$ is the lag of time $t$. As $y(t)$ has a domain of $[0,t]$, $|\tau|\leq t$. Since the autocorrelation function $R_y(\tau)$ is relevant with time $t$, the stock price $y(t)$ is a non-stationary stochastic process. Figure (\ref{fig:my_label_1}) illustrates $R_y(\tau)$.
\begin{figure}[htbp]
    \centering

\tikzset{every picture/.style={line width=0.75pt}} %set default line width to 0.75pt        

\begin{tikzpicture}[x=0.75pt,y=0.75pt,yscale=-1,xscale=1]
%uncomment if require: \path (0,300); %set diagram left start at 0, and has height of 300

%Straight Lines [id:da6023875592988355] 
\draw    (136,172.33) -- (510.33,172) ;
\draw [shift={(512.33,172)}, rotate = 179.95] [color={rgb, 255:red, 0; green, 0; blue, 0 }  ][line width=0.75]    (10.93,-3.29) .. controls (6.95,-1.4) and (3.31,-0.3) .. (0,0) .. controls (3.31,0.3) and (6.95,1.4) .. (10.93,3.29)   ;
%Straight Lines [id:da7557344441773295] 
\draw    (324.17,172.17) -- (324.33,6.33) ;
\draw [shift={(324.33,4.33)}, rotate = 90.06] [color={rgb, 255:red, 0; green, 0; blue, 0 }  ][line width=0.75]    (10.93,-3.29) .. controls (6.95,-1.4) and (3.31,-0.3) .. (0,0) .. controls (3.31,0.3) and (6.95,1.4) .. (10.93,3.29)   ;
%Straight Lines [id:da050666890780235674] 
\draw    (182.33,172) -- (324.33,59) ;
%Straight Lines [id:da9281414901857346] 
\draw    (480.33,172) -- (324.33,59) ;

% Text Node
\draw (172,177) node [anchor=north west][inner sep=0.75pt]   [align=left] {$\displaystyle -t$};
% Text Node
\draw (475,176) node [anchor=north west][inner sep=0.75pt]   [align=left] {$\displaystyle t$};
% Text Node
\draw (520,161) node [anchor=north west][inner sep=0.75pt]   [align=left] {$\displaystyle \tau $};
% Text Node
\draw (290,47) node [anchor=north west][inner sep=0.75pt]   [align=left] {$\displaystyle N_{0} t$};
% Text Node
\draw (336,7) node [anchor=north west][inner sep=0.75pt]   [align=left] {$\displaystyle R_{y}( \tau )$};

\end{tikzpicture}

    \caption{The Autocorrelation Function of Stock Price}
    \label{fig:my_label_1}
\end{figure}

In figure (\ref{fig:my_label_1}), $R_y(\tau)$ has a wide distribution, which means that $y(t)$ changes slowly over time and exists to have a large inertia or correlation. It indicates that there are laws that can be identified and utilized in stock price fluctuations, which are predictable. \cite{zhuang2001correlation} empirically analyze the autocorrelation function based on the Shanghai and Shenzhen CSI index from 19/12/1990 to 1/6/2000 and find a similar result as in figure (\ref{fig:my_label_1}).

The autocorrelation function essentially describes a certain dependence between the historical data of the stock price $y(t)$ and the future data $y(t+\tau)$, which is illustrated by equation (\ref{eq:9}). In other words, the historical data can be used to predict the future data. However, the correlation between $y(t+\tau)$ and $y(t)$ decreases linearly to zero with increasing horizons $\tau$.

\subsection{Stock Price Displacement Formula}
Suppose that in physics, stock price $y(t)$ is regarded as the displacement of the particle in the time interval $[0,t]$, then the average speed of $y(t)$ in the interval $[0,t]$ is given by:
                \begin{eqnarray}\label{eq:10}
				\overline{v(t)} = \frac{1}{t} \int_0^t x(t)dt .
			\end{eqnarray}
Thus, we can rewrite equation (\ref{eq:3}) as:
                \begin{eqnarray}\label{eq:11}
				y(t) = \left(\frac{1}{t} \int_0^t x(t)dt\right)\cdot t = \overline{v(t)}\cdot t .
			\end{eqnarray}
The displacement of stock price $y(t)$ is equivalent to the product of average speed $\overline{v(t)}$ and the time $t$, that is, stock price has a positive relationship with time. $\overline{v(t)}$ is the arithmetic mean of the white noise sampled signal $x_k (t)$, and in physics, it represents the DC component in $x_k (t)$, which reflects the deterministic part of $x_k (t)$. As time $t$ increases, the range of fluctuation of $\overline{v(t)}$ will gradually decrease and $\overline{v(t)}$ will stabilize. Meanwhile, $y(t)$ increases linearly with time.

\section{The Characteristics of Frequency Domain}
Stock price volatility is a time-domain signal that changes over time, so it is intuitive, simple, and easy to understand and analyze the structural characteristics of stock prices in the time domain. However, time-domain analysis studies the volatility as a whole. It cannot reflect the intensity distribution of harmonic components of different frequencies (or cycles). Besides, it cannot distinguish the effect of harmonic components of different frequencies (or cycles) on the overall volatility. Therefore, it is impossible to reveal the characteristics and laws of stock price volatility effectively.

Frequency-domain analysis can fill the gap of a time-domain analysis. This analysis can provide a certain formula. For example, equation (\ref{eq:5}) expresses the PSD of white noise in the frequency domain. Some certain rules and characteristics hidden in stochastic events are relatively easy to reveal as the frequency-domain analysis decomposes the stock price volatility into harmonic components of different frequencies. Meanwhile, it studies the intensity distribution in the frequency domain to find out the main frequency components that generate stock price volatility. The analysis provides strong evidence for clarifying the internal mechanism of stock price fluctuations, trend forecasting and risk management.

Within the interval $[0,t]$, the average power of $y(t)$ is finite and the autocorrelation function $R_y (\tau)$ is absolutely integrable. According to the Wiener-Khinchin theorem, the PSD of $y(t)$, $S_y (\omega)$, is the Fourier transformation of its autocorrelation function, given the expression as:
\begin{eqnarray}\label{eq:12}
				 S_y (\omega) &=& \int_{-\infty}^{+\infty} R_y (\tau) e^{-j\omega \tau} \nonumber \\
                   \;          &=& N_0 \cdot \frac{\textit{sin}^2 (\omega T)}{\omega ^2} \nonumber \\
                   \;          &=& N_0 T^2 \textit{sinc}^2 (\omega T)
			\end{eqnarray}
where $\textit{sinc}$ denotes a Sinc function with expression as:
                \begin{eqnarray}
				\textit{sinc}(x) = \frac{\textit{sin}(x)}{x} , s.t.\; x\neq 0 \nonumber
			\end{eqnarray}

\begin{figure}[htbp]
    \centering
    \includegraphics[width=15cm]{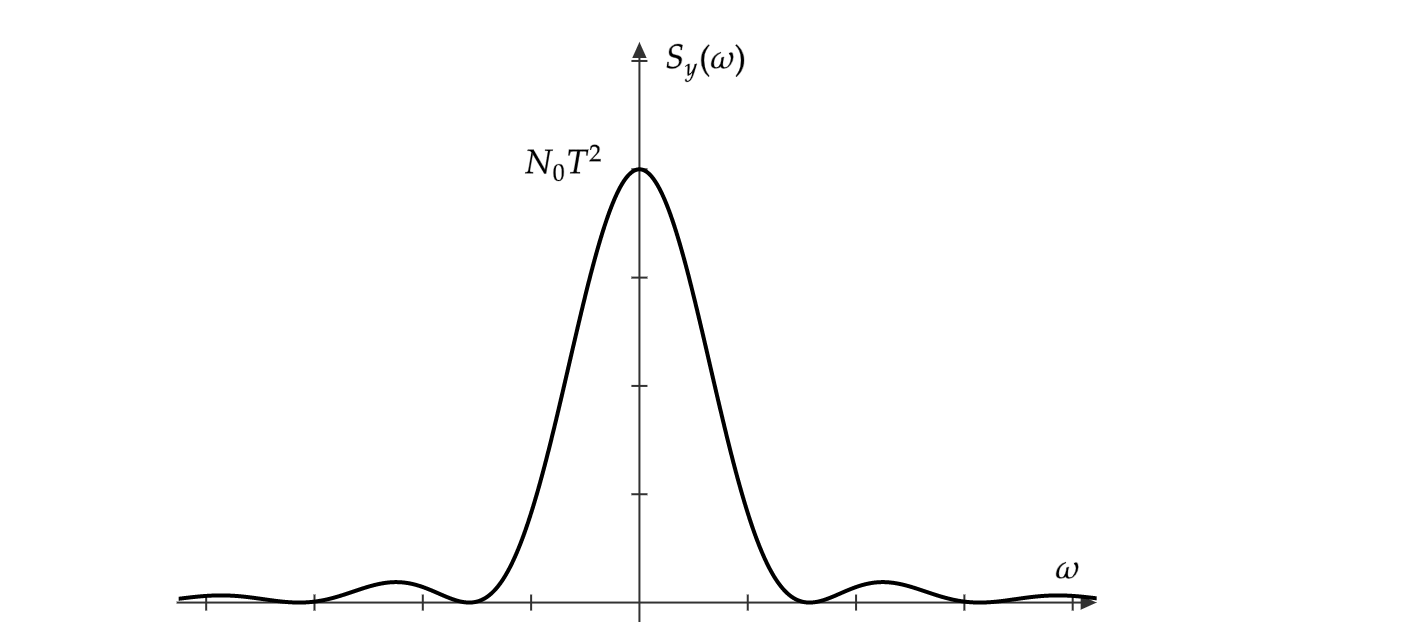}
    \caption{The PSD of Stock Price}
    \label{fig:my_label_2}
\end{figure}
Figure (\ref{fig:my_label_2}) illustrates $S_y (\omega)$. The frequency-domain characteristics of stock price are: (1) $S_y (\omega)$ is continuous with respect to $\omega$ and $y(t)$ is a nonperiodic signal in the time domain; (2) The harmonic amplitude of $y(t)$ is inversely proportional to the frequency $\omega$, indicating that the stock price is $1/f$ distributed and structurally invariant (self-similar) under the scale conversion; (3) $S_y (0)=N_0 T^2$ means that in the time domain, $y(t)$ has a linear trend line proportional with time $t$, and $y(t)$ fluctuates around the line; (4) The main lobe of the Sinc function ($-\pi/T \leq \omega \leq \pi/T$) concentrates more than 90\% of the fluctuation energy, so $y(t)$ is, in principle, pink noise, revealing the fact that the volatility of the stock price is mainly low-frequency, i.e. the movement has a large inertance (or correlation), ensuring the original trend and state under certain time and conditions.

\cite{andreadis2000self} calculates the PSD of the S\&P500 index from 1/12/1988 to 1/4/1998 and obtains the empirical result that the log-return of S\&P500 is inversely proportional to the squared frequency, corresponding with the implication in this paper.

\section{Conclusion}
This paper defines the quantitative relationship between stock prices and time as a certain time function. Given the basic law, “The first order condition of the log-return of the stock price is equal to white noise”, an integral white noise model is established to describe the stock price movement. This model reveals the characteristics and rules of stock price by deriving its autocorrelation function, displacement function and power spectral density (PSD). A theoretical framework has proved the predictability and long-run linear stock price trend, concluding that the PSD is inversely proportional to the squared frequency. This paper can correctly explain the past and present stock price movement and experiential facts. Meanwhile, it can describe and forecast the phenomenon, characteristics and rules of the future movement.

			\pagebreak
			\bibliographystyle{agsm}
			\bibliography{references}

\end{document}